\documentclass[a4paper,11pt]{article}
\usepackage{pos}
\usepackage{setspace}
\usepackage{hyperref}
\usepackage{floatrow}

\title{Radio astronomy locates the neutrino origin in bright blazars}
 \ShortTitle{Neutrinos from radio-bright blazars}

\author*[a,b]{Alexander Plavin}
\author[a,b,c]{Yuri Y. Kovalev}
\author[a]{Yuri A. Kovalev}
\author[d]{Sergey Troitsky}

\affiliation[a]{Astro Space Center of Lebedev Physical Institute,\\Profsoyuznaya 84/32, 117997 Moscow, Russia}
\affiliation[a]{Moscow Institute of Physics and Technology,\\Institutsky per. 9, Dolgoprudny 141700, Russia}
\affiliation[a]{Max-Planck-Institut f\"ur Radioastronomie,\\Auf dem H\"ugel 69, 53121 Bonn, Germany}
\affiliation[a]{Institute for Nuclear Research of the Russian Academy of Sciences,\\60th October Anniversary Prospect 7a, Moscow 117312, Russia}

\emailAdd{alexander@plav.in}

\abstract{High-energy astrophysical neutrinos have been observed by several telescopes in the last decade, but their sources still remained unknown. We address the problem of locating astrophysical neutrinos’ sources in a statistical manner. We show that blazars positionally associated with IceCube neutrino detections have stronger parsec-scale radio cores than the rest of the sample. The probability of a chance coincidence is only $4\cdot10^{-5}$ corresponding to a significance of $4.1\sigma$. We explicitly list four strong radio blazars as highly probable sources of neutrinos above 200 TeV: 3C 279, NRAO 530, PKS 1741$-$038, and PKS 2145+067. There are at least 70 more radio-bright blazars that emit neutrinos of lower energies starting from TeVs. Using continuous RATAN-600 monitoring of VLBI-selected blazars, we find that radio flares at frequencies above 10 GHz coincide with neutrino arrival dates. The most pronounced example of such behavior is PKS 1502+106 that experienced a major flare in 2019. We demonstrate that the majority of IceCube astrophysical neutrino flux derived from muon-track analyses may be explained by blazars, that is AGNs with bright Doppler-boosted jets. High-energy neutrinos can be produced in photohadronic interactions within parsec-scale relativistic jets. Radio-bright blazars associated with neutrino detections have very diverse gamma-ray properties, which suggests that gamma-rays and neutrinos may be produced in different regions of blazars and not directly related. A narrow jet viewing angle is, however, required to detect either of them. We conclude with discussion of recent independent tests and extensions of our findings.}

\FullConference{37$^{\rm{th}}$ International Cosmic Ray Conference (ICRC 2021)\\
		July 12th -- 23rd, 2021\\
		Online -- Berlin, Germany}

\begin{document}
\maketitle

\section{Introduction}

Astrophysical neutrinos of TeV energies have been convincingly detected by the IceCube experiment since 2012 \citep{IceCubeFirst26}. In 2019, the 
Baikal--GVD (Gigaton Volume Detector) experiment reported on the first few $E>100$~TeV neutrino candidates; indications of to the astrophysical high-energy neutrino flux were also found by the 
ANTARES experiment. Despite all these observations, the origin of energetic astrophysical neutrinos remained unknown until recently. Active galactic nuclei (AGNs) and blazars in particular have been considered a probable class of neutrino sources since the very early days of multimessenger astronomy. However, no significant connection between neutrino events and gamma-loud blazars has been found. This contrasts with the association of a single high-energy neutrino event with a gamma-ray flare in the blazar TXS~0506+056 and an excess of low-energy events from the same direction \citep{icecubecollaborationNeutrinoEmissionDirection2018}.

In this work based on \citep{neutradio1,neutradio2}, we report results of our search for neutrino associations using radio observations of blazars: very long baseline interferometry (VLBI) and long-term monitoring at an individual telescope, RATAN-600. VLBI observations provide an angular resolution high enough to distinguish central parsecs in AGNs even at cosmological distances. Activities seen by VLBI in the apparent jet base were shown to be good tracers of what is happening in and around the nucleus. Single-dish telescopes do not directly resolve inner regions of typical blazars; however, observed flares at scales of months and years probe parsec scales as well due to causality arguments. Radio observations are also used in ICRC~2021 proceedings searching for spatial and temporal associations of ANTARES and Baikal neutrinos with blazars: \citep{RadioPS,Illuminati2021,Baikal2021}.

\section{Data}
\label{s:data}

The largest sample of VLBI positions and flux densities of blazars is presented as the Radio Fundamental Catalog\footnote{\url{http://astrogeo.org/rfc/}} (RFC). A special effort was made to compile a complete sub-sample of AGNs limited by their VLBI flux density $S^\mathrm{VLBI}_\mathrm{8GHz}>150$~mJy at 8~GHz. This complete sample consists of 3411 objects. The resulting sky coverage is shown with gray dots in \autoref{f:skymap}. The RATAN-600 radio telescope of the Special Astrophysical Observatory (Russia) has been monitoring a sample of AGN selected on their VLBI flux density since late 1980s. For the analysis in this work, we use simultaneous measurements at 5, 8, 11, and 22~GHz.

\begin{figure*}
    \centering
    \includegraphics[width=\linewidth]{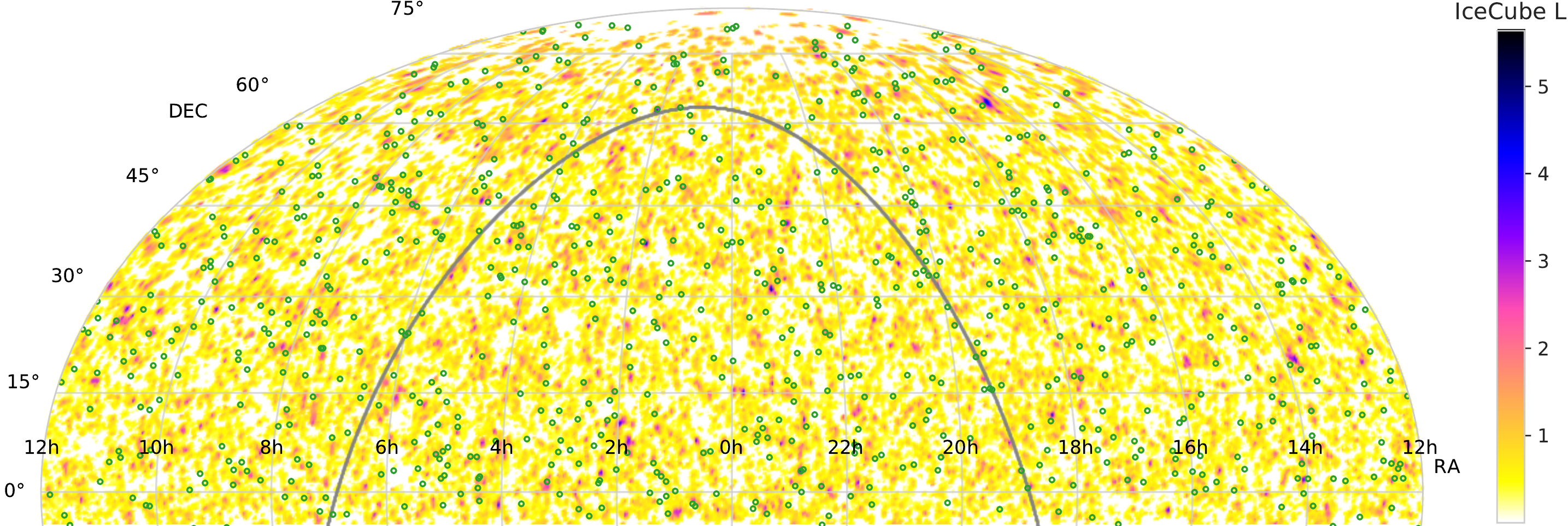}
    \caption{Sky map of the IceCube local $p$-value logarithms denoted as $L$. Darker areas with larger $L$ indicate higher probabilities to have an astrophysical neutrino point source in this direction. All sky north of $\delta = -5^\circ$ is displayed in equatorial coordinates. Radio AGNs from the complete 8~GHz VLBI sample down to 0.33~Jy are shown as green circles.}
    \label{fig:skymap_c}
\end{figure*}

We utilize neutrino detections at IceCube, with information presented as two separate datasets. The first contains events with the highest energies, above 200~TeV, that we compiled from a set of catalogs containing alerts and alert-like events. Overall, we select 57 track events from 2009 to 2019 with (i) energy $E\geq200$~TeV, and (ii) angular resolution $\Omega < 10$~sq.~deg. These events are visualized in \autoref{f:skymap}. The second IceCube dataset is based on the full sample of track events from 2008 to 2015 \citep{icecubecollaborationAllskyPointsourceIceCube2018}. It is published as pre-trial local $p$-values on a grid of pixels covering the entire sky, with events of $\sim10$~TeV energies dominating in this map. We treat the provided negative logarithms of $p$-values as a measure of the direction-dependent neutrino emission, and denote them as $L = -\log p$. Only the Northern sky at $\delta>-5^\circ$ is utilized in our work due to a heavily degraded IceCube sensitivity above the horizon. This area of the map is illustrated in \autoref{fig:skymap_c}.

\section{Detecting Neutrinos from Bright Blazars}
\label{s:results_bright}

\begin{figure*}[t]
    \centering
\includegraphics[width=1.0\textwidth]{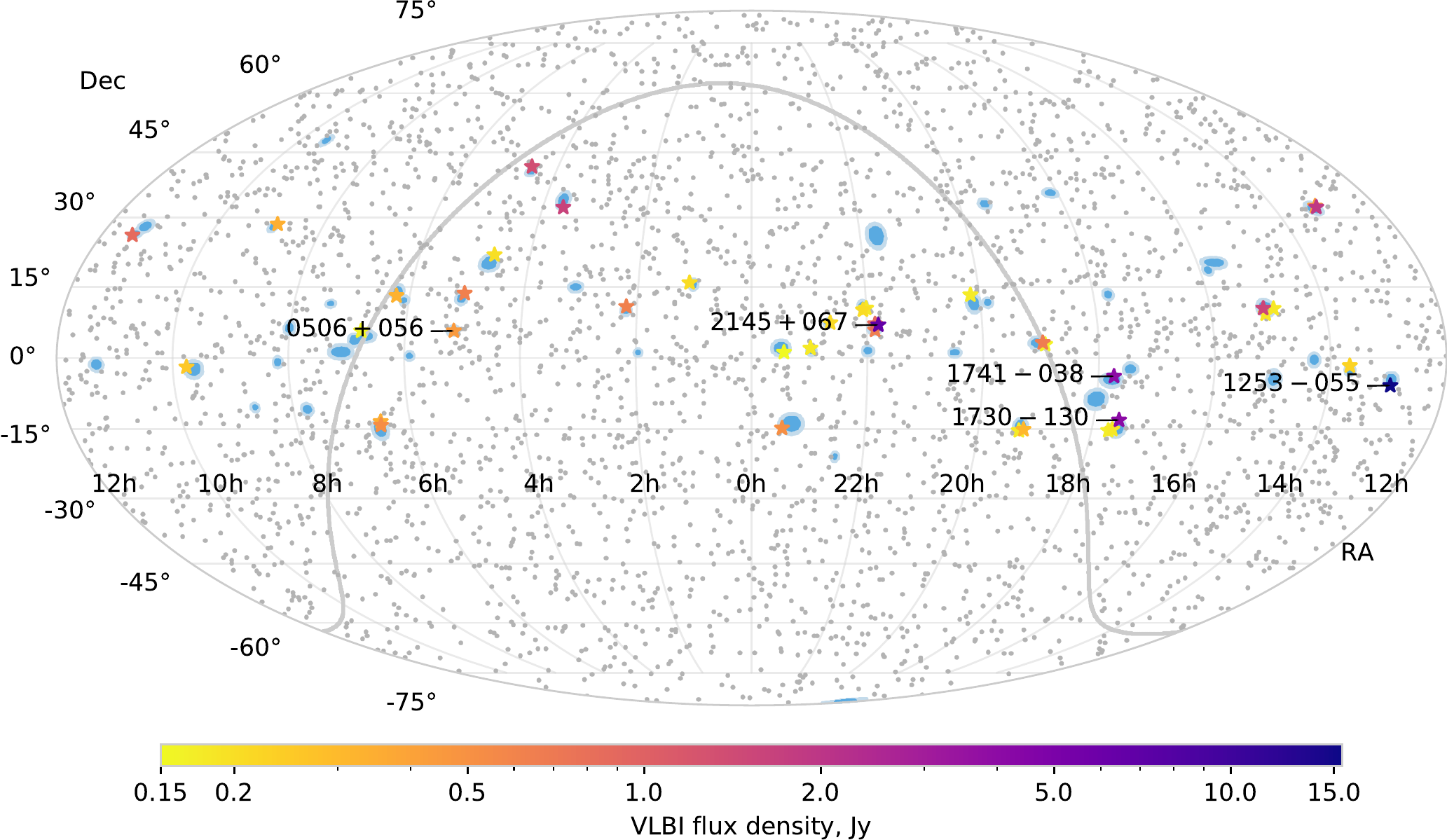}
\caption{
\label{f:skymap}
IceCube event locations on the sky, represented by blue ellipses. Stars represent all blazars from our complete VLBI sample within neutrino error regions. Other members of the VLBI sample are shown by gray dots.
The labeled objects denote four blazars with the strongest parsec-scales jets that are the most probable neutrino associations according to our analysis; we also show the location of the first neutrino association TXS~0506+056.}
\end{figure*}

\begin{figure}[t]
    \centering
    \floatbox[{\capbeside\thisfloatsetup{floatwidth=sidefil,capbesideposition={right,top},capbesidewidth=.45\linewidth}}]{figure}
    {\caption{Average of VLBI flux densities for AGN inside the IceCube error regions shown as a black triangle in comparison to 68\% Monte-Carlo interval (blue horizontal line) for randomly-shifted events. Flux densities for individual AGN inside the error regions are also shown as vertical black ticks for information.}
    \label{f:avg_fluxes}}
    {\includegraphics[width=\linewidth]{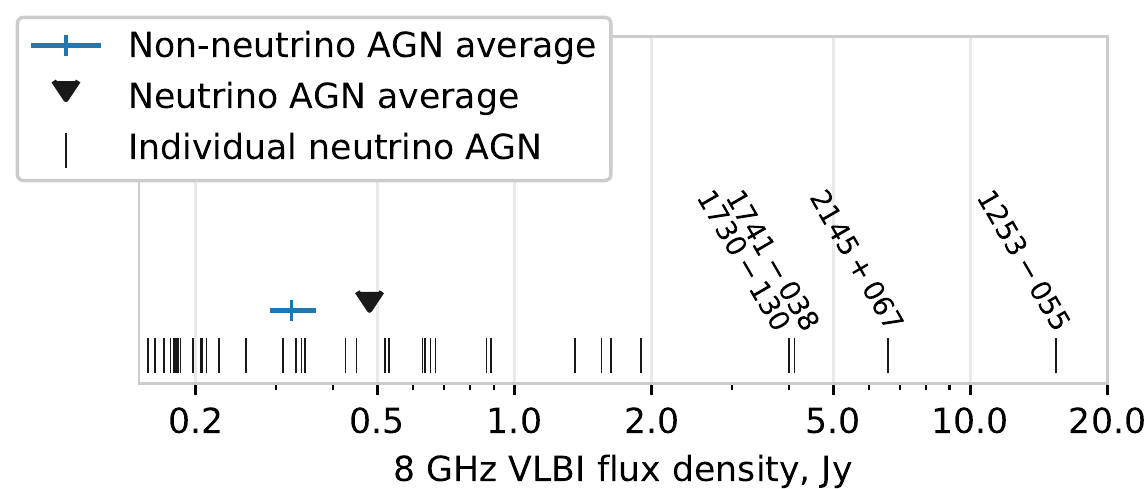}}
\end{figure}

\begin{figure}
    \centering
    \floatbox[{\capbeside\thisfloatsetup{floatwidth=sidefil,capbesideposition={left,top},capbesidewidth=.4\linewidth}}]{figure}
    {\caption{$P$-values of spatial blazar-neutrino correlation using the all-energies sky map; they are shown for a range of VLBI flux density cutoffs. Each value indicates the probability that the observed correlation between neutrinos and blazars stronger than the flux cutoff happened by a random chance. The lowest $p$-value of $4\cdot10^{-4}$ is attained for $S_{\rm min} = 0.33$~Jy.}
    \label{fig:logp_pval_flux}}
    {\includegraphics[width=\linewidth]{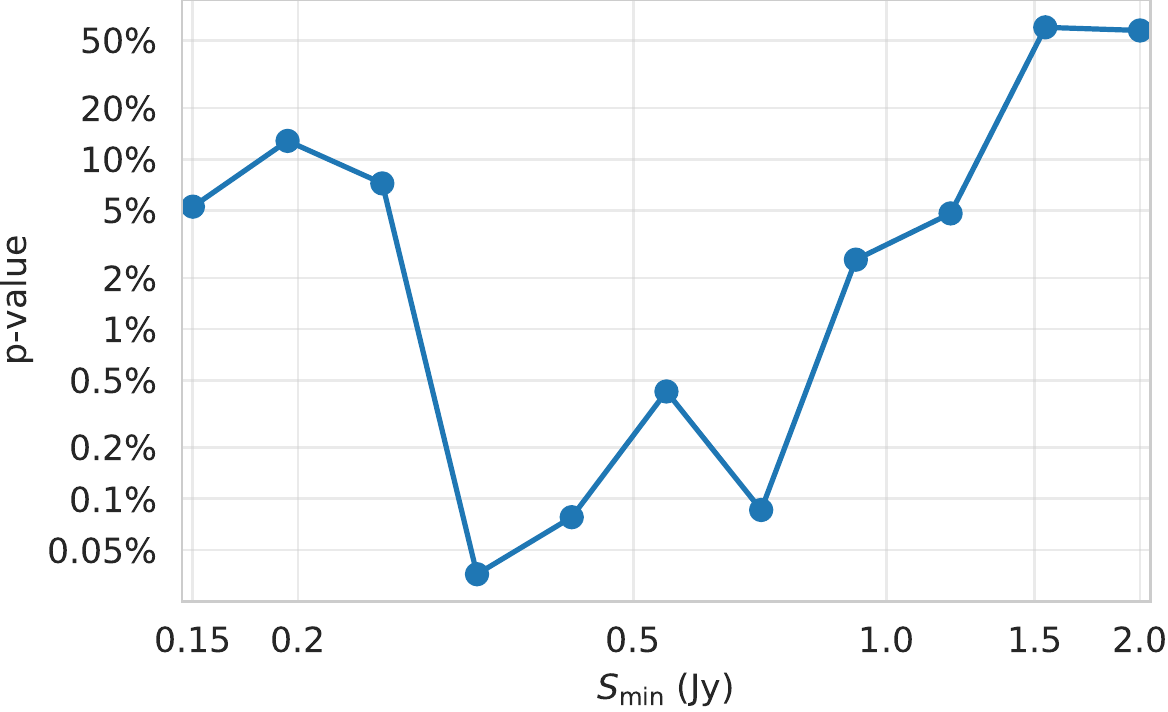}}
\end{figure}

We use the average VLBI flux density of blazars to evaluate a potential connection between radio emission from compact central regions and neutrino production. Specifically, we compute the average flux density over all blazars within error regions of IceCube events and take this value as the test statistic. Then, a Monte-Carlo method is employed to test if the average is significantly higher than could arise by chance. We obtain the distribution under the null hypothesis of no correlation by shifting each IceCube event to a random right ascension coordinate. Then the $p$-value is calculated as the fraction of random realizations with test statistic values greater or equal to the real one.

To implement this procedure, error regions for each track event have to be specified. We start with 90\% uncertainties reported in the catalog. Next, we need to account for systematic errors in IceCube event positions. They are not typically published but are always present, and related in particular to the lack of knowledge of ice properties; the IceCube upper limit estimate is $1.0^\circ$ \citep{IceCubeFirst26}. We choose to introduce the systematic error magnitude as a free parameter --- same for all events and directions on the sky --- and determine its optimal value. This is implemented by trying out multiple values of the systematic error (we take 11 values: $0^\circ, 0.1^\circ, \dots, 1^\circ$) to select the one yielding the strongest signal. The minimal of the 11 $p$-values represents the \emph{pre-trial} significance that is affected by the multiple comparisons issue. We obtain an unbiased \emph{post-trial} estimate via another Monte-Carlo procedure: all those steps are repeated for artificial samples to determine how often a lower $p$ arises in a chance fluctuation. This results in a post-trial $p$-value free from the multiple comparisons issue.

This approach results in the post-trial chance probability $p=0.07\%$ of the flux density of blazars around IceCube detections being as high as observed. We conclude that the neutrino-blazar spatial correlation is significant. The minimum pre-trial $p$-value is $0.03\%$ obtained for the additional error of $x = 0.5^\circ$. This $x$ can be interpreted as a rough estimate of IceCube systematic errors, which is in a good agreement with aforementioned IceCube upper limits, $<1^\circ$.

\autoref{f:skymap} demonstrates IceCube events on the sky together with blazars from our complete sample. \autoref{f:avg_fluxes} compares the average VLBI flux density for AGN within the neutrino error regions to Monte-Carlo realizations of this statistic for randomly-shifted positions of neutrino events. This figure highlights that the actual AGNs being selected as possible neutrino counterparts are, on average, very strong on parsec scales. We estimate how many sources drive the found correlation, and attempt to explicitly list them. For that, we repeat our analysis, dropping the strongest blazars one by one. The $p$-value rises above the 5\% level when four objects are removed, and we interpret this as a lower bound on the number of AGN likely emitting high-energy neutrinos. The four strongest blazars are 1253$-$055 (3C\,279), PKS~2145$+$067, PKS~1741$-$038, and 1730$-$130: none of them have been singled out as neutrino sources in the literature before.

We expand the found neutrino-blazar associations to lower energies using the IceCube sky map of neutrino likelihood described in \autoref{s:data}. Testing whether neutrinos are likely to arrive from bright blazars boils down to a comparison of $L$ values at blazar positions with $L$ values in other regions of the sky. We take the average $L$ at positions of blazars stronger than some flux density threshold $S_{\rm min}$ as the test statistic, and apply the Monte-Carlo procedure described above. Pre-trial $p$-values for each threshold value are shown in \autoref{fig:logp_pval_flux} with the minimum of $p=4\cdot10^{-4}$ attained for $S_{\rm min} = 0.33$~Jy. The resulting post-trial $p$-value is $3\cdot10^{-3}$ and represents the chance probability for $L$ at bright blazar positions to be as high as actually observed.

Having established that an excess of TeV neutrinos is detected from the directions of radio-bright blazars as well, we estimate how many objects drive this correlation. For a range of $L$ thresholds, we count blazars with $L$ value higher than the threshold. Then we subtract counts obtained in the same way for blazar positions randomly shifted in Right Ascension. The maximum excess count reaches $104 \pm 32$ close to the median $L$ threshold. This suggests that the majority of neutrino-emitting blazars have IceCube $L$ values that are not extremely high, but are close to the overall average. Such blazars can only be distinguished by a statistical approach, and would be lost in any analysis focused on the brightest regions of the IceCube map only.

The physical effects, objects, and their regions probed by our two analyses of higher- and lower-energy neutrino samples are effectively the same. Thus, we may combine the obtained statistical results after ensuring that they are independent; independence is achieved by dropping pixels within error regions of the high-energy events from the $L$ map. With these regions masked out, the post-trial $p$-value based on the $L$ map analysis marginally increases from $3\cdot10^{-3}$ to $4\cdot10^{-3}$. Finally, we follow the Fischer's method, and find the joint probability of a chance coincidence $p=4\cdot10^{-5}$ ($4.1\sigma$ for a normal distribution).

\section{Blazar Radio Flares and Neutrinos}

\begin{figure*}
    \centering
    \floatbox[{\capbeside\thisfloatsetup{floatwidth=sidefil,capbesideposition={right,top},capbesidewidth=.4\linewidth}}]{figure}
    {\caption{Average activity indices derived from RATAN-600 blazar monitoring observations, shown for a range of radio-neutrino delays. Filled areas correspond to curves of the same color and indicate pointwise 68\% null intervals based on Monte-Carlo realizations. \label{f:xcorr}}}
    {\includegraphics[width=1\linewidth]{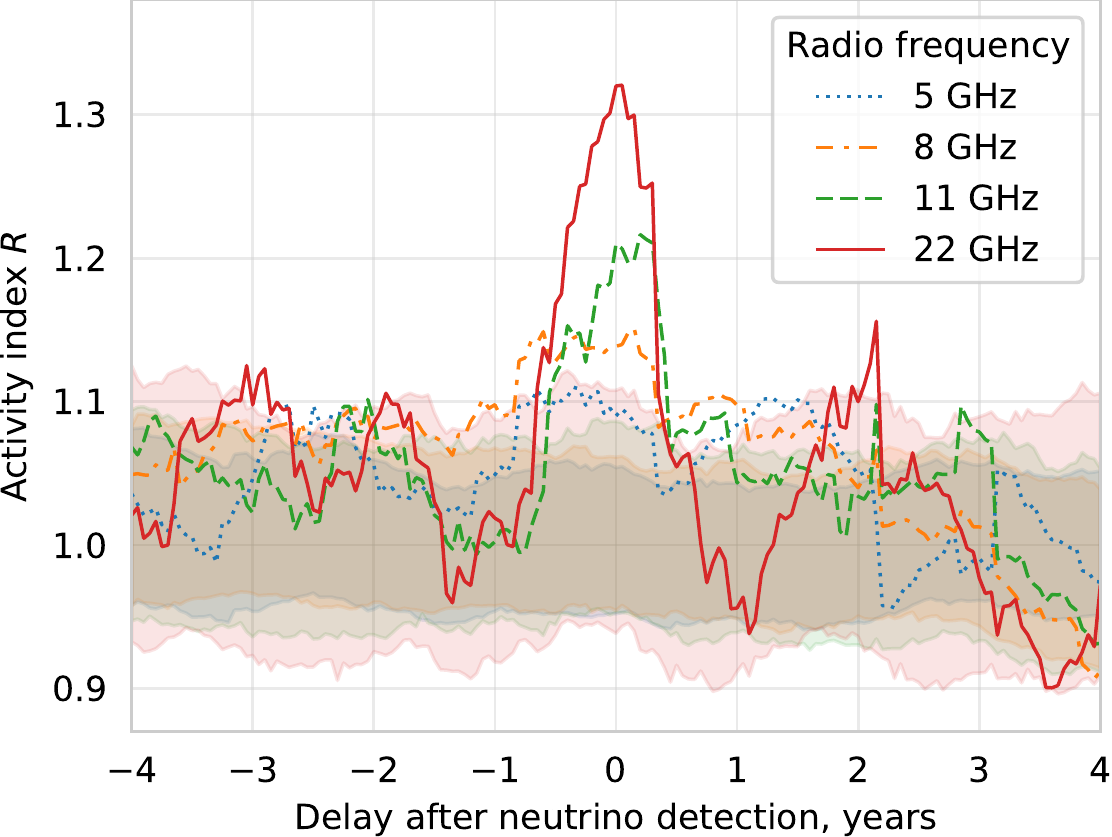}}
\end{figure*}

It was predicted \citep[see, e.g.,][]{Murase-rev} that neutrinos can be associated with flares in central regions of AGNs --- the immediate vicinity of the black hole or parts of the jet close to its origin. The studies of TXS 0506+056 \citep[e.g.,][]{IceCubeTXSgamma,r:kovalev0506} support this prediction, however it has not been established for larger samples of AGN. We evaluate the neutrino-flares correlation by analyzing radio flux density changes in temporal coincidence with IceCube neutrino detections. We focus on RATAN-600 observations at 22~GHz because flares are typically more pronounced and occur earlier at high frequencies due to synchrotron opacity effects in blazar jets.

For each blazar within the IceCube error regions, we compute the \textit{activity index} $R_{\text{22GHz}}^{t=0}$ defined as the ratio of average 22~GHz radio flux density within a $\Delta T$ time window around neutrino detection to the average value outside this time range. These ratios are averaged over all blazars to form a single number --- the test statistic. This value being higher than expected from statistical fluctuations would mean that neutrinos do correlate with flares seen in radio observations. We test this hypothesis with the same Monte-Carlo approach described in the previous section: plug the average $R_{\text{22GHz}}^{t=0}$ as the test statistic and use an additional trial range of $0.1\,\textrm{yr}\le\Delta T \le 2\,\textrm{yr}$ with 20 values spaced by $0.1\,\textrm{yr}$. The resulting post-trial $p$-value is 5\%; this is not strongly significant, but can definitely be considered suggestive given the established spatial correlation.

To visualize the correlation, we compute $R^t$ for all four frequencies and different time lags. These $R^t$ values are shown in \autoref{f:xcorr}. The plot indicates that at the highest frequencies, especially 22~GHz, there is a pronounced peak around zero delay. As mentioned above, the frequency trend is in an agreement with the synchrotron nature of jet emission. The detected correlation happens on timescales of months, which are the smallest scales that can be probed here due to the monitoring cadence. Furthermore, we repeat the flaring analysis after removing the four strongest blazars selected by our time-averaged analysis in the previous section. This does not affect the correlation strength noticeably. Thus, we conclude that the temporal correlation of the most energetic neutrinos with observed flares is present not only in those blazars distinguished by their time-averaged flux density. The highest individual activity index in our analysis belongs to the PKS~1502+106 blazar, which was noted to have a flare coinciding with a neutrino detection \citep{r:kielmann1502ATel}.

\section{Independent Tests and Extensions}
\label{s:tests}

Works that extend our findings or attempt to test them based on independent observations have started to appear recently. A temporal correlation of neutrinos with radio flares is found in \citep{2021A&A...650A..83H} based on monitoring at OVRO and Metsahovi observatories. Indications of blazar associations are seen in ANTARES and Baikal-GVD neutrino events as well, and preliminary results are reported in these proceedings \citep{Illuminati2021,RadioPS,Baikal2021}. A seeming exception is a work searching for the neutrino-blazar correlation based on the most recent 10-year IceCube data release \citep{2021arXiv210109836I}: they did not detect any significant effect \citep{2021PhRvD.103l3018Z}. Here we demonstrate that this result is in argeement with ours, and point out potential reasons for the lack of correlation.

In fact, \citep{2021PhRvD.103l3018Z} findings do not formally contradict our conclusions. We report in \citep{neutradio1}, that at least 25\% of astrophysical neutrino flux is associated with blazars in the VLBI sample, assuming the power law spectrum determined by IceCube muon-track diffuse search. Authors of \citep{2021PhRvD.103l3018Z} estimate that no more than 30\% of astrophysical neutrinos come from these blazars, which is consistent with 25\%. Nevertheless, it is instructive to understand why no detectable correlation shows up in the 10-year analysis. We list likely potential reasons below.

The analysis performed in \citep{2021PhRvD.103l3018Z} differs from both our approaches, and from the standard IceCube likelihood analysis. It is not a direct test of the same hypothesis on a larger sample, but an independent result and should be treated as such. Differences from the standard IceCube analysis partially stem from the lack of estimated neutrino energies in the 10-year event dataset: only energies of muons are present there. This makes properly accounting for spectral properties and different effective areas impossible. We use the aggregated 7-year sky map of IceCube likelihood in our analysis: this map is built using reconstructed neutrino energies and should give proper effective weights to events of different energies.

The IceCube PSF is assumed to be Gaussian in \citep{2021PhRvD.103l3018Z}, while its actual shape is different and has heavier tails \citep{2021arXiv210109836I}. The Gaussianity assumption strongly downweights events in those tails compared to either taking the proper PSF shape into account, or assigning equal weight to all matches within a certain distance. The former case is indicative of our study based on 7-year map in \citep{neutradio2}; the latter corresponds to the per-event analysis in \citep{neutradio1}. These PSF effects become even more important given the presence of systematic IceCube uncertainties. Systematic uncertainties are not reported in the released event catalogs, but are always present and are on the order of $0.5^\circ$ at hundreds of TeVs energies: see \autoref{s:results_bright}.

In the 10-year IceCube dataset, the reconstruction of event arrival directions and their uncertainties differs from previous datasets. Specifically, we find that some high-energy events utilized in our study \citep{neutradio1} have their positions and reported errors significantly changed. Differences are acknowledged in the IceCube dataset description \citep{2021arXiv210109836I}. They lead to major significance reductions in other analyses as well, including a factor of 100 higher $p$-value for the TXS~0506+056 flare in 2014. We follow the IceCube guidance accompanying the release: no problems have been identified with previously published datasets, there is no a priori reason to prefer the new sample over the old one, and thus the older results do not get superseded.

\section{Conclusion}

In this study, we establish that astrophysical neutrinos with energies from TeV to PeV are produced by numerous bright blazars, that is active galactic nuclei with jets pointed towards us. Comparison of the VLBI blazar sample with IceCube neutrino tracks yields the post-trial significance of the directional association of $4.1\sigma$: the probability of a chance coincidence is $p=4\cdot10^{-5}$. There are more than 70 radio-bright blazars that emit neutrinos of these energies. Moreover, we find a temporal correlation of neutrino arrivals with radio flares at frequencies above 10~GHz observed by the RATAN-600 telescope. The most pronounced example of this correlation is PKS~1502+106 that experienced a major flare in 2019. Independent extensions and tests of our observational findings have recently appeared. Associations are found based on OVRO and Metsahovi radio monitoring \citep{2021A&A...650A..83H}, on ANTARES and Baikal-GVD neutrino detection \citep{Illuminati2021,RadioPS,Baikal2021}. A search for the neutrino-blazar correlation in the 10-year IceCube data release not result in a significant effect \citep{2021PhRvD.103l3018Z}; see discussion of differences in \autoref{s:tests}.

Both the association with VLBI blazars and temporal correlation on scales of months and years indicate that neutrinos are emitted from central parsec-scale regions of active galactic nuclei. Their emission occurs predominantly along the jet direction due to beaming effects. In \citep{neutradio1,neutradio2} we further evaluate astrophysical models and conclude that ultrarelativistic protons up to $10^{16}$~eV and X-ray photons of keV energies are required to produce observed neutrinos. Estimates based on our observational results indicate that there are enough blazars in the sky to explain the majority of the astrophysical neutrino flux derived from IceCube tracks. Radio-bright blazars associated with neutrino detections have very diverse gamma-ray properties, suggesting that gamma-rays and neutrinos may be produced in different regions of blazars and not directly related. A narrow jet viewing angle is, however, required to detect either electromagnetic emission or neutrinos.

\small
\setstretch{0.8}
\bibliographystyle{ICRC}
\bibliography{references}

\end{document}